\documentclass{iaus}
\usepackage{graphicx}

\title[JD 11.~~Substructure of the bright points] %% give here short title %%
{Substructure of Quiet Sun Bright Points}

\author[Aleksandra Andic, Jongchul Chae \& P.R. Goode]   %% give here short author list %%
{Aleksandra Andic$^1$
 \and Jongchul Chae$^2$ \and Phillip R. Goode$^1$}

\affiliation{$^1$BBSO, 40386 N. Shore Ln.\\Big Bear City, CA-92314, USA\\ email: {\tt aandic@bbso.njit.edu} \\[\affilskip]
$^2$Astronomy Program, Department of Physics and Astronomy, Seoul National University \\ Seoul 151-741, Korea \\email: {\tt jcchae@snu.ac.kr}}

\pubyear{2010}
\volume{273}  %% insert here IAU Symposium No.
\pagerange{1--4}
\jname{Physics of Sun and Star Spots}
\editors{A.C. Editor, B.D. Editor \& C.E. Editor, eds.}
\begin{document}

\maketitle

\begin{abstract}
Since photospheric bright points (BPs) were first observed, there has been a question as to how are they structured. Are they just single flux tubes or a bundle of the flux-tubes? Surface photometry of the quiet Sun (QS) has achieved resolution close to $0.1''$ with the New Solar Telescope at Big Bear Solar Observatory. This resolution allowed us to detect a richer spectrum of BPs in the QS. The smallest BPs we observed with TiO 705.68 nm were $0.13''$, and we were able to resolve individual components in some of the BPs clusters and ribbons observed in the QS, showing that they are composed of the individual BPs. Average size of observed BPs was $0.22''$.
\end{abstract}

\firstsection % if your document starts with a section,
\section{Introduction}

Observations of the solar photosphere, so far, revealed a plethora of tiny bright features, usually concentrated in active regions or bordering the supergranules in the quiet Sun (QS). Near disk center, they appear as "Bright Points" (BPs), roundish or elongated bright features located in the intergranular dark lanes Dunn \& Zirker (1973), Mehltretter (1974), and Title {\it et al.} (1987).\par
So far, the areas chosen for the study of BPs were active regions or plage because with prior resolution, the plethora of BPs was visible only there. The achieved resolution of the New Solar Telescope (NST) at Big Bear Solar Observatory (BBSO) revealed to us  large number of BPs structures in the QS. The smallest observed BPs  were around $0.1''$ leaving their substructure still a puzzling subject. Considering the recent result that BPs are possible source of the acoustic oscillations (Andic {\it et al.} 2010) the importance of their study increases.This work presents a details that might contribute to the answer of the possible substructure of the BP.

\section{Data}

The data set used here was obtained on 29 July 2009 using the 1.6 m clear aperture New Solar Telescope (NST) at Big Bear Solar Observatory (Goode {\it et al.} 2010). The optical setup at the Nasmyth focus of the NST contained as its main components a TiO broadband filter centered at 705.68 nm and a PCO.2000 camera. Both are described in detail in Andic {\it et al.}
(2010). We observed the quietest Sun at the center of the solar disk.\par

The data sequence consists of 120 bursts with 100 images in each burst. Individual frames have an exposure time of $10$ ms and the cadence, between bursts, is $15$ s. The images have a sampling of $0.037''$ per pixel.\par

The images were speckle reconstructed based on the speckle masking method of von der L\"{u}he (1993) using the procedure and code described in W\"{o}ger {\it et al.} (2008). The cadence of our data provided us with a Nyquist frequency of $67$ mHz. After speckle reconstruction, the images were aligned using a Fourier routine. This routine uses cross-correlation techniques and squared mean absolute deviations to provide sub-pixel alignment accuracy. However, we did not implement sub-pixel image shifting to avoid the substantial interpolation errors that sometimes accompany the use of this technique. \par
To accurately measure the size of the structures, we first located the maximum intensity of the structure in time and space within our time series and used full-width at half maximum (FWHM) of a spatial profile that contained that point. Due to the irregularity in the shape of the structures, we usually chose the profile that encompasses the longest dimension of the structure.

\section{Results}

\begin{figure}[h]
% \vspace*{-2.0 cm}
\begin{center}
 \includegraphics[width=3.4in]{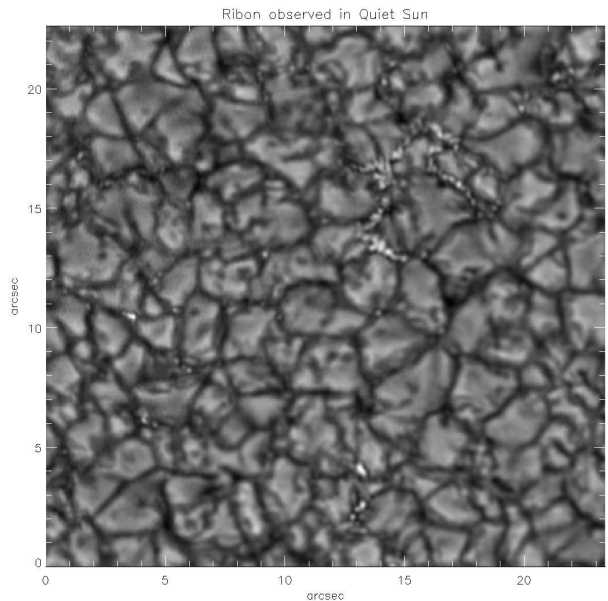} 
% \vspace*{-1.0 cm}
 \caption{The image shows a formation consisting of individual BPs which were previously seen in plague regions. This formation is called a ribbon. This ribbon is observed in the quietest sun, that is in the quiet sun area during the deep solar minimum.  Our resolution also made it possible to see the ribbon is composed of the individual BPs, which agrees with previous results.}
   \label{fig1}
\end{center}
\end{figure}

We observed the quiet Sun area during the deep solar minimum. In this area we found  the small scale structures that are typical for the active regions (Fig.\,\ref{fig1}).

\begin{figure}[h]
% \vspace*{-2.0 cm}
\begin{center}
 \includegraphics[width=5.4in]{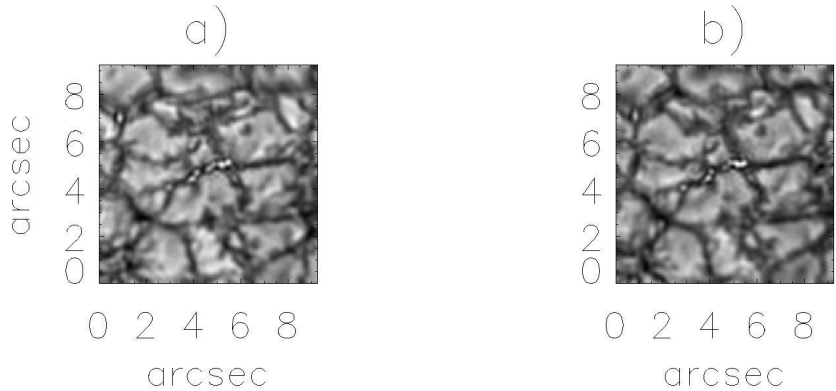} 
% \vspace*{-1.0 cm}
 \caption{The image shows two different stages in the BPs evolution. Left panel shows 6 BPs in one lane, all individually resolved. The right panel shows the situation around 1 min later where 3 of BP s are merged in one structure. This is typical behavior when we have, at least, 2 BPs close together. }
   \label{fig2}
\end{center}
\end{figure}

The BPs analyzed formed the structures called clusters by Viticchie {\it et al.} (2009) terminology. The BPs tended to form persistent groups that stayed on the same location for duration of our time series.  BPs in those groups tend to be attracted to each other until they reach a proximity that we cannot resolve, so they form a single structure, a cluster. An example of the process is visible at Fig.\,\ref{fig2} where we can see in the left panel one large elongated BP, while right panel shows the same structure 90 sec later after structure split into 2 separate BPs.

The larger clusters consisted of 3 BPs that were individually resolved at some point during our observations. Those clusters  had the elongated shapes with the longest dimensions over $1''$ in length. Most of the observed elongated shapes were clusters of BPs that during our time series unveiled their components, consisting of 2 or 3 separate BPs. Size of those clusters is varying from $0.48''$ to $1.2''$. The lifetime of the individual clusters varied from $0.7$ min to $30$ min. \par
\begin{figure}[h]
% \vspace*{-2.0 cm}
\begin{center}
 \includegraphics[width=5.4in]{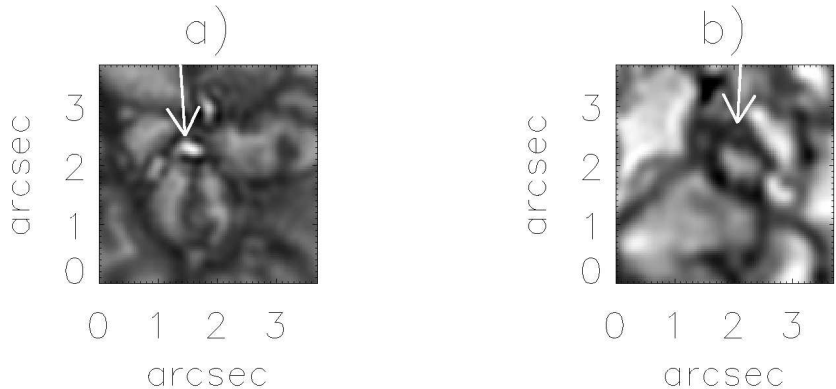} 
% \vspace*{-1.0 cm}
 \caption{The largest and the smallest registered bright point are shown. Panel a) shows one of the large BP-like structures (arrow points to it) across the length of it measures $0.9''$.  Panel b) shows the smallest BP we detected. It's diameter measures $0.13''$, at the limit of our achieved resolution indicating the possibility that the true structure might be even smaller.}
   \label{fig3}
\end{center}
\end{figure}

The average size of the observed BPs was $0.22''$, within a range from $0.13''$ for the smallest observed structure to $0.48''$ for the largest individual structure (Fig.\,\ref{fig3}). Right panel of Fig.\,\ref{fig3} shows one of the smallest observed structure. Another typicality is that those structures tended to have rather low intensity. Considering the facts that their size was at the limit of our achieved diffraction limit and their low intensity, we can speculate that those structures are smaller than the resolution power of our telescope.\par 
Their average lifetime was $4.28$ min, in the range of our imposed minimum of $45$ s to the longest lifetime of $29.75$ min. There is no clear relationship between the size and the lifetime of the analyzed BPs.\par

\section{Conclusions}

Starting with the ribbons that we detected in the quietest sun, we observed constant substructure. Even individual BPs that seem as a compact elements turned to be composed from one or more substructures that revealed themselves in course of the time. This happened for most of the BPs that were at least $0.5"$ in diameter. 
Our observation of BPs in the groups showed the tendency of the BPs to drift toward each other until they form a single structure. This result is similar to the observations made by Viticchie {\it et al.} (2009) in G-band line. Previously, it was reported that BPs could have various shapes (Dunn \& Zirker 1973, Mehltretter 1974, Title {\it et al.} 1987).
However, most of the elongated shapes we observed were clusters of BPs, which during our time series unveiled their components. \par 
All BPs that had diameter close to the $0.13"$ showed low intensity when compared with the larger BPs. Since their size was matching our achieved diffraction limit we might speculate that in reality they are much smaller. And that their intensity is low because it fills our resolution element. \par
To complement this result a signature of the flux-tubes collision inside a single bright point that did not separate itself has to be mentioned. Details concerning this finding are presented at the poster by A. Andic presented at this same conference.
All this results encourage us to speculate that BPs, as we observe them now, are composed of the multiple flux tubes.

P.R.Goode and NST are supported by following grands: NSF (AGS-0745744), NASA (NNX08BA22G) and AFOSR (FA9550-09-1-0655).

\begin{discussion}

No discussion was recorded. 

\end{discussion}

\end{document}